\documentclass[preprint,10pt]{elsarticle}
\usepackage{graphics}
\usepackage{mathrsfs}
\usepackage[percent]{overpic}
\usepackage{amsmath}
\usepackage{setspace}
\usepackage{natbib}
\usepackage[margin=2.5cm]{geometry}
\begin{document}

\title{Alpha decay as a probe for the structure of neutron-deficient nuclei}%

\author{Chong Qi}%
\ead{chongq@kth.se}
\address{Department of Physics, Royal Institute of Technology (KTH), SE-10691 Stockholm, Sweden}

\begin{abstract}
The advent of radioactive ion beam facilities and new detector technologies have opened up new possibilities to investigate the radioactive decays of highly unstable nuclei, in particular the proton emission, $\alpha$ decay and heavy cluster decays from neutron-deficient (or proton-rich) nuclei around the proton drip line. It turns out that these decay measurements can serve as a unique probe for studying the structure of the nuclei involved. On the theoretical side, the development in nuclear many-body theories and supercomputing facilities have also made it possible to simulate the nuclear clusterization and decays from a microscopic and consistent perspective. In this article we would like to review the current status of these structure and decay studies in heavy nuclei, regarding both experimental and theoretical opportunities. We then discuss in detail the recent progress in our understanding of the nuclear $\alpha$ formation probabilities in heavy nuclei and their indication on the underlying nuclear structure.
\end{abstract}

\maketitle

\section{Introduction}
There has been a long history of studies on the $\alpha$ radioactivity which was first described by Ernest Rutherford in 1899. The structure of the particle was identified  by 1907 as $^4$He (He$^{2^+}$ ion) with two protons and two neutrons, which, with the binding energy 7.1 MeV per nucleon, is the most stable configuration below $^{12}$C. 
The greatest challenge then was to understand how the $\alpha$ particle could leave the less stable mother
nucleus without any external disturbance.
The decay process was successfully interpreted  by Gamow
\cite{Gamow1928} and Gurney and Condon \cite{Gurney1928,PhysRev.33.127} as a quantum tunneling effect, which required to accept the
probabilistic interpretation of Quantum Mechanics. The extent to which this was
revolutionary can perhaps best be gauged by noticing the multitude of
models that have been put forward as an alternative
to the probabilistic interpretation. Besides its pioneering role in nuclear physics and in the development of quantum theory,
the tunneling effect is also realized to be responsible for the thermonuclear reactions and stellar evolution.
Processes like nuclear fusion, proton and $\alpha$ captures can also be explained as an inverse tunneling \cite{RevModPhys.70.77}.
The tunneling was accepted as a general physical phenomenon around  mid-20th century and also  becomes relevant at the nanoscale with important applications such as the tunnel
diode, scanning tunneling microscopy and quantum computing as well as chemical and biological
evolutions. Without tunneling there would be no star, no life, let alone nuclear physics or quantum mechanics.

\begin{figure}
\begin{center}

\begin{overpic}[width=1.1\linewidth]{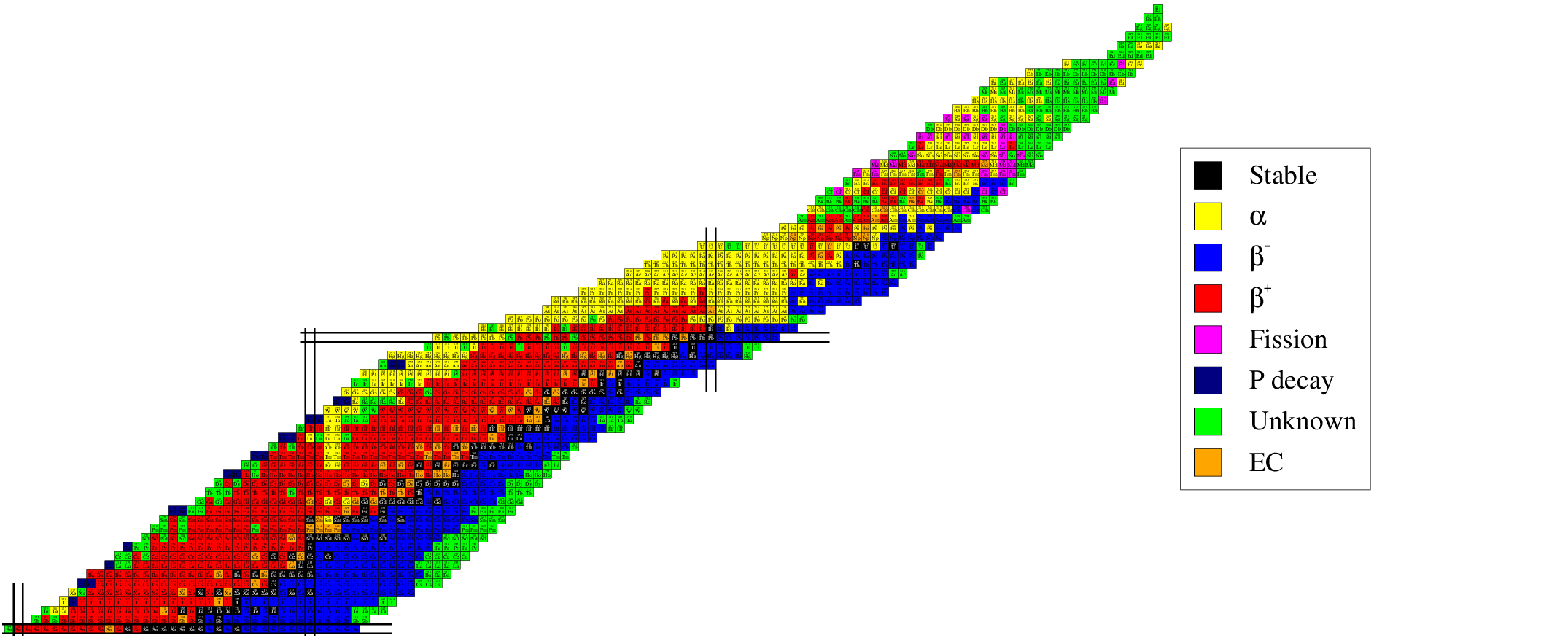}
\put(0,15){\Large Z}
\put(40,0){\Large N}
\end{overpic}

\end{center}
\caption{Dominant ground-state decay modes for nuclei with proton number $Z\geq50$. EC and P stand for electron capture and proton radioactivity, respectively. The horizontal and vertical lines correspond to the proton shell closures $Z=50$, 82 and neutron shell-closures $N=50$, 82 and 126, respectively. The shell structure in the superheavy nuclei is not known hitherto. \label{fig1}}
\end{figure}

$\alpha$ decay has been among the most important decay modes
of atomic nuclei for more than a century. The decay occurs most often in massive nuclei that have large proton to neutron ratios, where it can reduce the ratio of protons to neutrons in the parent nucleus, bringing it to a more stable configuration in the daughter nucleus. Almost all observed proton-rich or neutron-deficient nuclei
starting from mass number $A \sim 150$ have $\alpha$
radioactivities, as shown in Fig. \ref{fig1}. 
Various phenomenological and microscopic models
have been developed to study the $\alpha$-decay process, which can successfully reproduce available experimental $\alpha$-decay half-lives.
The spontaneous emission of charged
fragments heavier than the $\alpha$ particle is known as cluster
radioactivity. This process is more closely related to spontaneous fission, i.e., a disintegration of the heavy nucleus into two lighter ones \cite{0034-4885-62-4-001,RevModPhys.85.1541,2015arXiv151107517S}.
For available superheavy elements or superheavy nuclei \cite{0954-3899-34-4-R01,PT.3.2880,0954-3899-42-11-114001}, fission and $\alpha$ decay are the dominant decay modes. The detection of emitted $\alpha$ particles has been the principal method of identifying superheavy nuclei as well as their excited states \cite{PhysRevLett.111.112502}, which can be created in heavy ion fusion reactions.

Nuclear physics is undergoing a renaissance with the availability of intense radioactive beams. The new facilities have opened up new possibilities to investigate highly unstable nuclei as well as to probe existing formalisms trying to describe those nuclei. Recent investments in new or upgraded facilities such as FAIR at GSI, Darmstadt, HIE-ISOLDE at CERN, Geneva, SPIRAL2 at GANIL, Caen, FRIB at MSU and RIBF at RIKEN, in conjunction with new detector systems, in particular $\gamma$ ray tracking devices like AGATA, will produce unprecedented data on exotic nuclei and nuclear matter in the decades to come.  In this review we would like to discuss the recent developments and new opportunities in the study of the decay of heavy nuclei and our understanding of the so-called nuclear $\alpha$ formation probabilities and the underlying structure of the nuclei involved.
We will concentrate in particular on the progress that has been made during the past decade and the  current status of experimental and theoretical studies. Extensive reviews on the $\alpha$ clustering in light nuclei, which is a closely related topic, could be found, e.g., in Refs. \cite{Freer2014,Beck2014,vonOertzen200643}.

\section{The microscopic description of $\alpha$ decay}
The Gamow theory explained nicely the $\alpha$ decay as the penetration (tunneling) through the Coulomb barrier.
Although successful,  one can
assert that this is an effective theory, where one has to assume a preformed $\alpha$ particle inside the nucleus and concepts like ``frequency of
escape attempts" have to be introduced.  This semiclassical picture collides with basic quantum
mechanics, since even if the $\alpha$ particle existed in the mother nucleus, the Pauli
principle would hinder any free motion of the particle inside the nucleus. 
Actually it has been realized in the early study of nuclear structure that the nucleus cannot be composed of $\alpha$ particles \cite{RevModPhys.8.82}. The $\alpha$ configuration is usually a very small component of the nuclear wave function.
What is missing in Gamow's picture
is the probability that the
$\alpha$ particle is formed at a certain distance around the nuclear surface.
A proper
calculation of the decay process needs to address first the formation of the $\alpha$ particle
 around the nuclear surface and, in a
second step, the evaluation of the penetrability (the probability of tunneling) through the static Coulomb
and centrifugal barriers
at the region where
the $\alpha$ particle was already formed. 
It is
expected that the decays of the proton and other charged
clusters heavier than $\alpha$ can be described by the same
mechanism.

We understand now that the structure of the nucleus is best described by the nuclear shell model where its building blocks, neutrons and protons, are held together by an
average potential (the so-called nuclear mean field) generated by nucleon-nucleon potentials.
The shell structure indicates that nucleons need to fill successively single-particle orbitals separated by the magic numbers.
The traditional ones are 2, 8, 20, 28, 50, 82, and 126. 
The neutrons and protons are expected to distribute homogeneously if no two-body correlation is considered. However, they do
correlate with each other through the residual interaction (in the sense that the mean field part is subtracted from the nucleon-nucleon interaction).
The correlation may induce clusterization (i.e., large spatial overlap) of the four nucleons which may eventually become the $\alpha$-particle.
A proper description of $\alpha$ clusterization in terms of its
components
requires the treatment of the residual correlation in a microscopic many-body framework that is a challenging
undertaking.  Moreover, the nuclear shell structure may evolve as a function of isospin (or neutron/proton ratio), leading to different particle correlation properties.   
One of the main aims of modern nuclear structure studies is to address on the same footing the underlying nature
of atomic nucleus and the limit to its existence. It is also hoped that one can describe simultaneously dynamical processes including nuclear decay, reaction and fission.
The so-called \textit{ab initio} approaches (in the sense that the full realistic nucleon-nucleon interaction is used) have been developed in recent years and are able to describe light nuclei with $A<16$ with the help of supercomputing facilities. 
Because of the enormous configuration spaces involved, the properties of intermediate-mass nuclei are best described by the nuclear configuration interaction approach (the modern shell model) where one considers only the residual correlation between particles around the surface. The
superfluid nuclear density functional theory (e.g., the self-consistent Hartree-Fock-Bogoliubov method) provides a convenient tool to study the ground states and low-lying quasiparticle states of heavy nuclei throughout the nuclear chart.
 The main ingredient of such approach is an effective density-dependent two-body interaction that generates the nuclear mean field on top of which the pairing correlation is added.

\subsection{The $\alpha$ formation probability}
A variety of theoretical models were proposed for the explanation of the $\alpha$ decay
phenomenon~\cite{Lovas1998,Delion2010book,qi-alpha} (see, also, Ref. \cite{Mang1964} for a review on early efforts). 
Here we very briefly go through the microscopic $R$-matrix description of the $\alpha$ decay \cite{RevModPhys.30.257,Thomas01091954} for which details may be found in recent publications \cite{Delion2006,PhysRevC.61.024304,PhysRevC.46.1346,PhysRevC.44.545,PhysRevC.86.034338,Qi2009,QI2009a,Qi2012c,Qi2010c}. In general, the $\alpha$-decay half-life can be 
written as 
\begin{equation}\label{life}
T_{1/2}=\frac{\hbar\ln2}{\Gamma_{\alpha}} \approx \frac{\ln2}{\nu} \left|
\frac{H_l^+(\chi,\rho)}{RF_{\alpha}(R)} \right|^2,
\end{equation}
where $\nu$ is the velocity of the emitted $\alpha$ particle with angular momentum $l$ which is equal to zero for ground-state to ground-state $\alpha$ decays of even-even nuclei. 
The quantity $F_{\alpha}(R)$ is the formation
amplitude of the $\alpha$ cluster at distance $R$. 
$R$ is usually chosen at a distance around the
nuclear surface where the internal wave function $F_{\alpha}(R)$ is matched with the
wave function of the outgoing $\alpha$ particle. 
$H_l^+$ is the Coulomb-Hankel function with $\rho=\mu\nu R/\hbar$ and
$\chi = 4Z_de^2/\hbar\nu$. $\mu$ is the reduced mass 
and $Z_d$ is the charge number of the daughter
nucleus. 
The penetrability
is proportional to $|H_l^+(\chi,\rho)|^{-2}$. 
Its great importance in
radioactive decay studies lies in the fact that within a given
decay the penetrability process is overwhelmingly dominant. 
The amplitude of the wave
function inside the nucleus is defined as
\begin{equation}\label{foram}
F_{\alpha}(R)=\int d{\mathbf R} d\xi_d d\xi_{\alpha}
[\Psi_d(\xi_d)\phi(\xi_{\alpha})Y_l(\mathbf R)]^*_{J_mM_m}
\Psi_m(\xi_d,\xi_{\alpha},\mathbf{R}),
\end{equation}
where $d$, ${\alpha}$ and $m$ label the daughter, emitting $\alpha$ and mother
nuclei, respectively, and $\xi$ denote the coordinates of the nucleons involved.  $\Psi_m$ and $\Psi_d$  are the wave functions of the mother and daughter nuclei. $\phi(\xi_{\alpha})$ is a Gaussian function of the relative
coordinates of the nucleons that constitute the $\alpha$ particle. 

We take $^{212}$Po as a simple example. 
The nucleus can be described as a four-particle state ($\alpha_4$) outside the doubly magic $^{208}$Pb (with frozen degrees of freedom).
 The wave function can be
written within the shell model framework as 
\begin{equation}
\label{msmwf}
|^{212}{\rm Po}(\alpha_4)\rangle=\sum_{\alpha_2 \beta_2} X(\alpha_2\beta_2;\alpha_4) 
|^{210}{\rm Pb}(\alpha_2)\otimes^{210}{\rm Po}(\beta_2)\rangle
\end{equation} 
where $\alpha_2$ ($\beta_2$) labels two-neutron (two-proton) states. 
The amplitudes $X$ are
influenced by the neutron-proton (np) interaction. If this interaction was neglected, 
only one configuration would appear in Eq.~(\ref{msmwf}). This
is in cases where the correlated four-particle
state is assumed to be provided by collective vibrational states. As a result,
calculations can be performed by assuming $|^{212}{\rm Po(gs)}\rangle$ 
as a double pairing vibration above the $^{208}$Pb inert core, i.e.,
$
|^{212}{\rm Po(gs)}\rangle=|^{210}{\rm Pb(gs)}\otimes~^{210}{\rm Po(gs)}\rangle$.
The corresponding formation amplitude acquires the form
\begin{eqnarray}\label{po212}
{F}_{\alpha}(R;^{212}{\rm Po(gs)}) =
\int d\mathbf{R}
d\xi_{\alpha}\phi_{\alpha}(\xi_{\alpha})
\Psi_{2\nu}(\mathbf{r_1}\mathbf{r_2};^{210}{\rm Pb(gs)})
\Psi_{2\pi}(\mathbf{r_3}\mathbf{r_4};^{210}{\rm Po(gs)}),
\end{eqnarray}
where $\mathbf{r_1},\mathbf{r_2}$ ($\mathbf{r_3},\mathbf{r_4}$)
are the neutron (proton) coordinates and $\mathbf{R}$ is the center
of mass of the $\alpha$ particle.
If we assume that the intrinsic wave function of the
$\alpha$ particle can be approximated by a $\delta$ function, an even
simpler expression exists for the $\alpha$ formation amplitude, which reads,
\begin{equation}\label{delta}
F_{\alpha}(R) = \frac{16\pi^2}{R^4}\left(\frac{
s_{\alpha}^3}{3}\right)^{3/2}\Psi_{2\nu}(R,R,0)\Psi_{2\pi}(R,R,0),
\end{equation}
where $s_{\alpha}=\sqrt{20}/3R_{\alpha}$, $R_{\alpha}\sim 1.281$ fm is the root mean square radius of the $\alpha$ particle and we take
$\mathbf{\hat{r}_1} =\mathbf{\hat{r}_3} =\mathbf{\hat{z}}$. This approximation works well outside the nuclear surface.

In the first applications of the shell
model to the description of the mother nucleus of $\alpha$ decay only one 
configuration was used. The results were discouraging 
since the theoretical decay rates were smaller than the corresponding
experimental values by 4-5 
orders of magnitude \cite{Mang1964,PhysRev.119.1069,SANDULESCU1962332,Mang57,Fliessbach197675,PhysRevC.13.1318}, depending on the value to
be chosen for the nuclear radius.  However, since the matching radius $R$ has to be chosen at a
distance beyond the point where the cluster was formed, i. e.  
beyond the range of the nuclear force 
and Pauli exchanges, 
the formation amplitude had to be or should have been evaluated at rather 
large distances. But that 
would have required shell model calculations  with 
large bases for the mother and daughter nuclei. With the 
very limited shell-model spaces used at that time, 
the region of prominent four-particle correlation was not reached.
The fundamental role of configuration mixing was confirmed by actual 
large-scale calculations \cite{TONOZUKA197945,PhysRevC.27.896}.  This surface $\alpha$-clustering effect produces a tremendous enhancement of the $\alpha$-decay widths in both $^{212}$Po \cite{TONOZUKA197945} and light nuclei \cite{Arima1974475,ARIMA197215}. 
 With the expression for the formation amplitude shown above, the experimental 
half-life can now be reproduced rather well if a large number of
high-lying configurations is included \cite{TONOZUKA197945}.  Recent calculations in Ref. \cite{Qi2012c} are done within the harmonic oscillator (HO) representation by using a surface delta interaction and
nine major HO shells.

\subsection{The single-particle basis and the Hartree-Fock wave function}

The evaluation of $\alpha$ formation amplitude involves the evaluation of the overlap between the corresponding proton and neutron radial functions in the laboratory framework with
the $\alpha$-particle intrinsic wave function as defined in the center of mass framework (see, e.g., Ref. \cite{PhysRevC.69.044318}). The transformation can be relatively easily handled if the radial wave functions are defined within the harmonic oscillator basis due to its intrinsic simplicity. This is also the reason why the harmonic oscillator representation is used in most \textit{ab initio} and shell-model configuration interaction calculations.
More realistic calculations are done based on Woods-Saxon and Nilsson single-particle states. A single particle basis consisting of two different harmonic oscillator representations was introduced in Ref. \cite{PhysRevC.54.292}.
An additional attractive pocket potential of a Gaussian form was introduced on top of the Woods-Saxon potential in  Ref. \cite{PhysRevC.87.041302} in order to correct the asymptotic behavior of the  $\alpha$ formation amplitude. The mixture of shell model and cluster wave functions was considered in Ref. \cite{VARGA1992421} and was applied to describe the decay of the ground state of $^{212}$Po. The calculated formation probability that can reproduce experimental decay half-life is found to be 0.025.
 
Significant progress has also been made in the development of nuclear density functional approaches which are now able to provide a reasonable description
of ground state binding energies
and densities throughout the nuclear chart, even though the description of the single-particle spectroscopy is still less satisfactory. The Skyrme-Hartree-Fock single-particle wave functions were applied to calculate the $\alpha$ formation amplitudes in both even-even nuclei
\cite{PhysRevC.88.064316,1402-4896-89-5-054027} and even-odd nuclei \cite{PhysRevC.92.014314}. However, the calculated formation amplitude is still several of orders of magnitude too small in comparison to experimental data. The application of the recently refined functional seems to make the discrepancy even worse \cite{1402-4896-89-5-054027}. Further investigation along this line would be interesting to understand the origin of the discrepancy, which may shed additional light on the constraint of the density functional.
Time-dependent Hartree-Fock calculations for $\alpha$ decay and $\alpha$ capture were carried out in Ref. \cite{San83} with a simplified Skyrme plus Yukawa potential. No spin-orbital field was considered.

\subsection{Continuum effect}
A full microscopic description of the clustering on the nuclear surface requires the use of realistic finite single-particle
potentials including their continuum states, which is still a challenging open problem. The continuum is expected to be important since the decay involves states at the nuclear surface and high-lying states beyond that.
The influence of the single-particle resonances on $\alpha$ clustering was considered in Refs. \cite{PhysRevC.48.1463,PhysRevC.61.024304}. 
In Refs. \cite{PhysRevC.86.034338,1742-6596-436-1-012061} the complex-energy shell model was applied to describe the $\alpha$ decay of $^{212}$Po and $^{104}$Te by using a  simple separable interaction. The single-particle space is again  expanded in a Woods-Saxon basis that consists of bound and unbound resonant states. 
The calculations for $^{104}$Te did not fully converge in that work, which is probably due to the fact that the valence proton
shells that lie in the continuum were not considered in the model space.

\subsection{Nuclear deformation}
Eq.~(\ref{life}) is
valid for the decays of spherical as well as deformed nuclei~\cite{Lovas1998}. If the Coulomb barrier of a deformed nucleus is also deformed (or with anisotropic barrier width or height), the tunneling of the $\alpha$ particle  may become direction dependent.
The tunneling through a deformed Coulomb barrier was first described within the WKB approximation by Bohr, Fr\"oman, and Mottelson, who introduced the so-called Fr\"oman matrix. The method was applied in later calculations on $\alpha$ decay \cite{PhysRevC.69.044318,PhysRev.103.1298,PhysRev.181.1697,PhysRev.112.512}.
The angular distribution of emitted $\alpha$ particles from deformed nuclei  were measured in Refs.
\cite{PhysRev.124.1512,PhysRevC.2.2379,PhysRevC.71.044324,PhysRevLett.77.36}, which indeed revealed preferential $\alpha$ emission  along the symmetry axis. However, it should be mentioned that the anisotropy can be a combined effect of nuclear deformation and structure \cite{PhysRevLett.82.4787}.

Semi-classical approaches were also proposed to treat the nuclear deformation in a macroscopic way 
\cite{PhysRevC.72.064613}. This was also used recently in Refs.
\cite{PhysRevC.92.014602,Qian201487}. Coupled-channel calculations were presented in Ref. \cite{PhysRevC.92.051301} and compared with that from the averaged WKB approach.

\section{$\alpha$ formation probability and pairing correlation}
The mechanism for nuclear pairing is similar to that behind electronic superconductivity \cite{bohr1998nuclear2}.
The nuclear pairing correlation is related to the presence of strongly attractive two-body pairing interaction with angular momentum $J=0$. 
It is the most crucial correlation beyond the nuclear mean field and leads to  zero angular momentum (e.g., with all particles paired to $J=0$) for the ground states of all observed even-even nuclei. 
It is also responsible for the occurrence of systematic staggerings, depending on the evenness and oddness of $Z$ and $N$,
 in many nuclear phenomena including the nuclear binding energy. The pairing correlation is relatively less favored in nuclei with odd numbers of protons and/or neutrons in relation to the fact that the odd neutron and/or proton do not participate the pairing correlation.

The pairing wave function can be described well by the Barden-Cooper-Schrieffer (BCS) approach. More sophisticated models have also been developed in recent years.
The pairing correlation manifests itself through the coherent contribution of a large number of shell-model configurations. 
This feature is  also responsible
for the two-particle clustering, which is
manifested in a strong increase in the form factor of the
two-particle transfer cross section in
transfer reactions between collective pairing states.
This also gives rise
to giant pairing resonances, which correspond to the most collective of
the pairing states lying 
high in the spectrum. 
Soon after 
the pairing interaction had been adapted 
to nuclei, it was applied to the study of $\alpha$ decay \cite{Soloviev1962202}. 
The pairing correlation 
highly enhances the 
calculated $\alpha$-decay width and is indeed  the mechanism governing the formation of $\alpha$ particles at the nuclear surface.

\begin{figure}
\begin{center}
\includegraphics[scale=0.55]{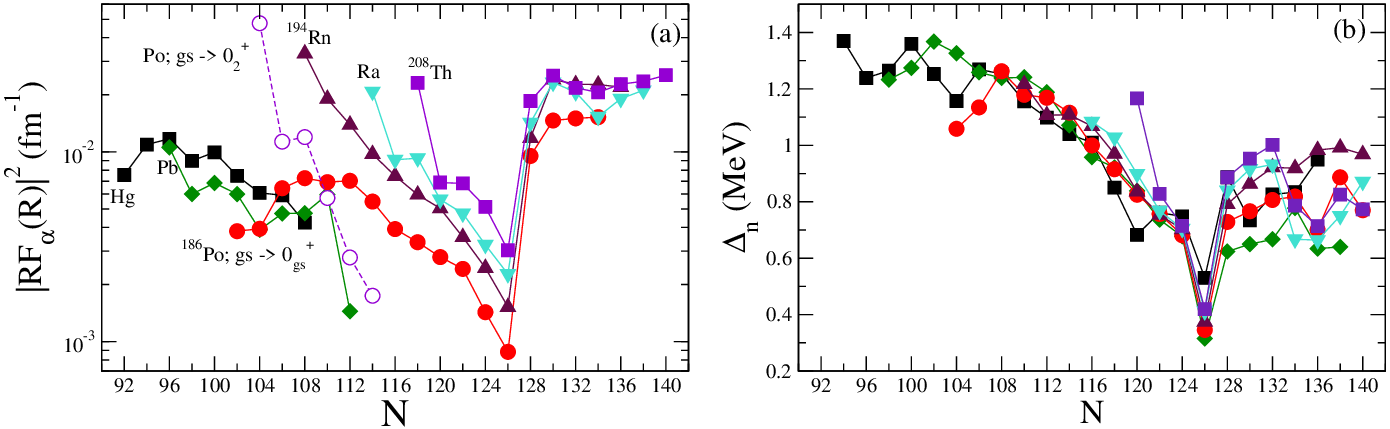}
\end{center}
\vspace{-0.8cm}\caption{(color online).  Upper panel: $\alpha$-particle
formation probabilities for the decays of the even-even isotopes 
as a function of the neutron numbers $N$ of the mother nuclei. 
 Lower panel: Neutron pairing gaps in even-even lead to thorium nuclei
extracted from experimental binding energies. From Ref. \cite{Andreyev2013}.
}
\label{fap}
\end{figure}

The formation
amplitude $F_{\alpha}(R)$ can be extracted from the experimental
half-lives $T^{{\rm
Expt.}}$ by
\begin{equation}
\log |RF_{\alpha}(R)|=\frac{1}{2}\log \left[ \frac{\ln
2}{\nu}|H^+_0(\chi,\rho)|^2\right] - \frac{1}{2}\log T^{{\rm
Expt.}}_{1/2}.
\end{equation}
This is done in Refs. \cite{Qi2010c,Andreyev2013,Qi2014203}. 
Fig. \ref{fap}a shows the formation probabilities $|RF_{\alpha}(R)|^2$ extracted from the experimental half-lives
from known
ground state to ground state $\alpha$-decay transitions in even-even isotopes from $N=92$ to 140. From the
trend of $|RF_{\alpha}(R)|^2$ around the neutron shell closure at $N=126$, one can deduce a global trend. Below
the shell closure, $|RF_{\alpha}(R)|^2$ decreases as a function of
rising neutron number, reaching its lowest values at the shell closure. When
the shell closure is crossed, a sudden increase in $|RF_{\alpha}(R)|^2$  is observed. It is followed by an additional but smaller increase and finally saturation occurs.
 The  $\alpha$-particle formation amplitudes for nuclei $^{162}$W, $^{162}$Hf \cite{PhysRevC.92.014326} and $^{193}$At \cite{Ket2003} are systematically larger than those of neighboring nuclei, which is not understood and needs further investigation.

Within the BCS approach the two-particle formation amplitude is proportional
to $\sum_k u_kv_k$ where $u_k$ and $v_k$ are the standard occupation numbers. To this
one has to add the overlaps of the corresponding proton and neutron radial functions with
the $\alpha$-particle  wave function on the nuclear surface,
which do not differ strongly from each other for neighboring nuclei. The BCS pairing gap is given by
$\Delta=G\sum_k u_kv_k$,
where $G$ is the pairing strength. It indicates that
the $\alpha$ formation amplitude is proportional to the product of the proton and the neutron pairing gaps which can serve as a signature of the
change in clusterization
as a function of the nucleon numbers. To probe this conjecture one may compare the
formation probabilities extracted from the experimental half-lives to the
corresponding pairing gaps. The latter can readily
be obtained from the experimental binding energies as \cite{Andreyev2013,Satula1998,Changizi2015210,PhysRevC.91.024305,Qi2012g}
\begin{equation}\label{dexp}
\Delta_n(Z,N)=\frac{1}{2}\left[B(Z,N)+B(Z,N-2)-2B(Z,N-1)\right].
\end{equation}
The empirical pairing gaps are shown as a function of the neutron number in Fig. \ref{fap}b.
One indeed sees a striking similarity between the tendency of the pairing gaps
in this figure with the $\alpha$-particle formation probabilities.
This similarity makes it possible to draw conclusions on the
tendencies of the formation probabilities. The near constant value of $|RF_{\alpha}(R)|^2$ for neutron
numbers $N\leq114$ is due to the influence of the $i_{13/2}$ and other high-$j$ orbitals. As these highly degenerate shells
are being filled the pairing gap and
the formation probability
should remain constant, as indeed they do in Fig. \ref{fap}.
A quite sharp decrease of 
formation probability and pairing gap happens as soon as the low-$j$ orbitals like $2p_{3/2}$, $1f_{7/2}$ and $2p_{1/2}$ start
to be filled. Finally, when we reach $N=126$, the pairing gap reaches its lowest value.
The possible influence of the $Z=82$ shell closure on the $\alpha$ formation probability and the robustness of the shell was also discussed in Ref. \cite{Andreyev2013}, which was questioned based on earlier measurements on the $\alpha$ decays of neutron-deficient Pb isotopes \cite{PhysRevC.60.011302}.
The role of the pairing interaction in multi-quasiparticle isomeric states and  the reduction of pairing in those states on $\alpha$-decay half-lives was examined in Ref. \cite{PhysRevC.90.044324}.

\subsection{Generic form of the $\alpha$ formation probability}
A generic form for the $\alpha$-particle formation probability was proposed in Refs. \cite{Andreyev2013,Qi2014203}: When the nucleons are filling a new major closed  shell, the $\alpha$-particle formation amplitude is nearly constant as high-$j$ orbitals are filled first. As soon as the low-$j$ orbitals are filled, the formation probability smoothly reduces until one reaches again a closed proton or neutron configuration. Crossing the closed shell induces a steep increase and the approximately constant trend mentioned above continues. However, when strong particle-hole excitations across closed shells are encountered, this 'generic' form of the $\alpha$-particle formation probability is altered as one clearly sees in the light polonium isotopes.

\subsection{$\alpha$ decays to and from excited states}
The $\alpha$ decays from ground states to  excited states (fine structure) as well as the decays from excited states  are usually less favored than ground-state to ground-state decays.
Ref. \cite{PhysRev.113.1593} first estimated the ratio between reduced widths for transitions to
ground and excited states. Further calculations on the decay to vibrational states were done in Ref. \cite{Sandulescu1965404} and later in Refs. \cite{PhysRevC.71.044315,PhysRevC.75.054301,PhysRevC.64.064302}. Systematics to rotational states in deformed in nuclei was done in Refs. \cite{PhysRev.181.1697,PhysRevC.78.034608,PhysRevC.73.014315,PhysRevC.81.064318}.
Systematic evaluations of the $\alpha$-decay fine structure were also done recently in Refs. \cite{PhysRevC.92.021303,Delion20151}.  It was found that the $\alpha$ decays to excited states also follow the Viola-Seaborg law, discussed in chapter 4.

The $\alpha$ decays of neutron-deficient nuclei around $Z=82$ are of particular interest in relation to the possible co-existence of states with different shapes \cite{RevModPhys.83.1467}. Three low-lying $0^+$ states in $^{186}$Pb were observed following the $\alpha$ decay of $^{190}$Po in Ref. \cite{10.1038/35013012}, which were interpreted to be of spherical, oblate and prolate shapes, respectively.
The $\alpha$ decay of $^{187}$Po to the spherical ground state  of $^{183}$Pb was observed to be strongly hindered \cite{PhysRevC.73.044324} whereas the decay to a low-lying excited state at 286 keV is favored. Based on the potential energy surface calculations, the $^{187}$Po ground state and the 286 keV excited state in $^{183}$Pb were interpreted as of prolate shape. The decay to the $^{183}$Pb ground state is hindered since this state has a spherical nuclear shape which is different from that of the ground state. The difference in the shapes indicates that the configurations of the mother and daughter wave functions would be very different. As a result, the $\alpha$ formation amplitude is significantly reduced. 
The hindrance of the $\alpha$ decay of the isomeric state in $^{191}$Po has the same origin \cite{PhysRevLett.82.1819}. The hindrance of the $\alpha$ decays of neutron-deficient even-even nuclei around $Z=82$ was measured in Ref. \cite{PhysRevLett.72.1329}.
The $\alpha$ decays to and from the excited $0^+_2$ states in Po, Hg and Rn isotopes were studied in Refs. \cite{Delion1995a,PhysRevC.54.1169,PhysRevC.90.061303}. These states are described as the minima in the potential energy surface  provided by the standard deformed Woods-Saxon potential. A simple approach was also presented in Ref. \cite{Karlgren2006} to evaluate the hindrance by taking the ratio between the wave function amplitudes for the transitions to the ground and excited $0^+$ states of the daughter nucleus obtained from potential energy surface calculations.

The robustness of the $N=Z=50$ shell closures has fundamental influence  on our understanding of the structure of nuclei around the presumed doubly magic nucleus $^{100}$Sn. 
It was argued that $^{100}$Sn may be a soft core in analogy to the soft $N=Z=28$ core $^{56}$Ni. It seems that such a possibility can be safely ruled out based on indirect information from recent measurements in this region \cite{PhysRevC.84.041306,Back2013,10.1038/nature11116,PhysRevLett.110.172501,PhysRevC.91.061304}. It is still difficult to measure the single-particle states outside the $^{100}$Sn core. The neutron single-particle states $d_{5/2}$ and $g_{7/2}$ orbitals  in $^{101}$Sn, which have been expected to be close to each other, 
were
observed by studying the $\alpha$-decay chain $^{109}$Xe$\rightarrow$ $^{105}$Te $\rightarrow$ $^{101}$Sn \cite{PhysRevLett.97.082501}. 
In Ref. \cite{Seweryniak2006}, the nucleus $^{105}$Te was also populated and one $\alpha$ transition was observed.
A prompt 171.7 keV $\gamma$-ray transition was observed in Ref. \cite{PhysRevLett.99.022504} and was interpreted as the transition from the $g_{7/2}$ to the $d_{5/2}$ orbital, which was assumed to be the ground state. 
On the other hand, two $\alpha$ decay events from $^{105}$Te were observed in Ref. \cite{PhysRevLett.105.162502} with the branching ratios (energies) of 89\%
(4711 keV) and 11\% (4880 keV). Based on those observation and on the assumption that the ground state of $^{105}$Te has spin-parity $5/2^+$, a flip between the $g_{7/2}$  and $d_{5/2}$ orbitals was suggested.
This information was used in the optimization of the effective shell-model Hamiltonian for this region \cite{PhysRevC.86.044323}. 

Excited states in the heavy nucleus $^{212}$Po were populated in Refs. \cite{Astier2010,PhysRevLett.104.042701} by using the $\alpha$ transfer reaction.  Several electric dipole (E1) transitions were observed, which are several orders of magnitude faster than one would expect between normal shell model states. The states involved were discussed in terms of enhanced $\alpha$ clustering structure. Those E1 transitions were evaluated in Ref. \cite{PhysRevC.85.064306} within the shell model approach by adding an additional Gaussian-like component in the single-particle orbitals to simulate the clustering. The enhanced experimental E1 strength distribution below 4 MeV in rare-earth nuclei was studied in Ref. \cite{PhysRevLett.114.192504} within the interacting boson model by treating the nucleon pairs as boson particles.

\section{The Geiger-Nuttall law and its generalizations}
The incredible range of $\alpha$ decay half-lives can be modeled with the so-called Geiger-Nuttall law~\cite{gn1,gn2}, where a striking correlation between the half-lives of radioactive
decay processes and the decay $Q_{\alpha}$ values (total amount of energy released by the decay process)
is found to be
\begin{equation}\label{gn-o}
\log T_{1/2}=\mathcal{A}Q_{\alpha}^{-1/2}+\mathcal{B},
\end{equation}
where $\mathcal{A}$ and $\mathcal{B}$ are constants that can be determined by fitting to experimental data.
The Gamow theory reproduced the Geiger-Nuttall law nicely by describing the $\alpha$ decay as the tunneling through the Coulomb barrier, which leads to the $Q_{\alpha}^{-1/2}$ dependence. Still one may wonder why the Geiger-Nuttall has been so successful. The reason is
that the $\alpha$-particle formation probability usually varies from
nucleus to nucleus much less than the penetrability. This is a consequence of the smooth variation in the nuclear structure that is often found when going from a nucleus to its neighbors.
In
the logarithm scale of the Geiger-Nuttall law, the differences in the
formation probabilities are usually small fluctuations (as seen in Fig. \ref{fap}) along the straight
lines predicted by that law~\cite{Buck90}. 

The Geiger-Nuttall law in
the form of Eq.~(\ref{gn-o}) has limited prediction power since its
coefficients change for the decays of each isotopic
series~\cite{Buck90}. Intensive works have
been done trying to generalize the Geiger-Nuttall law for a
universal description of all detected $\alpha$ decay
events \cite{PhysRevC.80.024310,0954-3899-39-1-015105}. One of the most known generalization is the  Viola-Seaborg law \cite{VIOLA1966741} which for even-even nuclei reads
\begin{equation}\label{gn-v}
\log T_{1/2}=(aZ_d+b)Q_{\alpha}^{-1/2}+bZ_d+d
\end{equation}
where $a$ to $d$ are constants and $Z_d$ the charge number of the daughter nucleus.

The importance of a proper treatment of $\alpha$ decay was attested in Refs. \cite{Qi2009,QI2009a} which shows that the different lines can be merged into a
single line. 
In this generalization the penetrability is
still a dominant quantity where $H^+_0(\chi,\rho)$ can be well
approximated by an analytic formula
\begin{equation}
H^+_0(\chi,\rho) \approx (\cot
\beta)^{1/2}\exp\left[\chi(\beta-\sin\beta\cos\beta)\right].
\end{equation}
 By defining
the quantities $\chi' = Z_{\alpha}Z_d\sqrt{A_{\alpha d}/Q_{\alpha}}$ and $\rho' =
\sqrt{A_{\alpha d}Z_{\alpha} Z_d(A_d^{1/3}+A_{\alpha}^{1/3})}$ where $A_{\alpha d}=A_d A_{\alpha}/(A_d+A_{\alpha})$,
one gets, after some simple algebra,
\begin{eqnarray}\label{gn-2}
\log T_{1/2}=a\chi' + b\rho' + c,
\end{eqnarray}
where $a$, $b$, $c$ are constants to be determined.

One thus obtained a generalization of the Geiger-Nuttall law
which holds for all isotopic chains and all cluster radioactivities.
The expression reproduces nicely most available
experimental $\alpha$ decay data on ground-state to ground-state radioactive decays. There is a case where it fails by a
large factor. This corresponds to the $\alpha$ decays of nuclei with 
neutron numbers equal to or just below $N=126$.
The reason for this large discrepancy is that 
the $\alpha$ formation amplitudes in $N\leq126$ nuclei 
are much smaller than the average quantity predicted. The case that shows the largest deviation corresponds to the $\alpha$ 
decay of the nucleus $^{210}$Po for which, as discussed in the previous section, the $\alpha$ formation is not favored due to the fact that the neutron states behave like holes below the shell closure.

\subsection{Limitations of the Geiger-Nuttall law}
The origin and physical meaning
of the coefficients $\mathcal{A}$ and $\mathcal{B}$ in the Geiger-Nuttall law can be deduced by comparing Eq. (\ref{gn-o}) and (\ref{gn-2}).
These coefficients are determined from experimental data and show a linear dependence upon $Z$. 
The need for a different linear $Z$ dependence of the coefficients $\mathcal{A}$ and $\mathcal{B}$ in different regions of the nuclear chart was discussed in Ref. \cite{Qi2014203}, which is related to the generic form of the $\alpha$ formation probability.
When the dependence of $log_{10}|RF_{\alpha}(R)|^2$ on the neutron 
number is not linear or constant, the Geiger-Nuttall law is broken.  This 
also explains why the Geiger-Nuttall law works so well for nearly all $\alpha$ emitters known 
today, as the data within each isotopic chain are limited to a region 
where $log_{10}|RF_{\alpha}(R)|^2$ is roughly a constant or behaves linearly with $N$.

For the polonium isotopic chain with $N<126$, the linear behavior of $log_{10}|RF_{\alpha}(R)|^2$ breaks down below $^{196}$Po.
As a result, the Geiger-Nuttall law is broken in the light polonium isotopes. This violation is induced by the strong suppression of the $\alpha$ formation 
probability due to the fact that the deformations (or shell-model configurations) of the ground states of the lightest $\alpha$-decaying
neutron-deficient polonium isotopes ($A < 196$)  are very different from those of the daughter lead isotopes.

\subsection{The effective approaches}
The simple Gamow theory is so
successful that even today it is applied, with minor changes, in the studies
of radioactive decays. That is, the $\alpha$ particle (or charged clusters in general) is
assumed to be a preformed particle which  is initially confined in a finite potential well, bouncing on and reflected off the internal wall of the
potential. The $\alpha$ particle (with no intrinsic structure) wave function is assumed to be an eigenstate of the potential for which the depth can be determined by fitting to the $Q_{\alpha}$ value according to the
Bohr-Sommerfeld quantization condition~\cite{Buck90}.
The decay
width is given as
\begin{equation}
\Gamma =F_{{\rm
eff}}\exp\left[-2\int^{R_2}_{R_1}k(r)dr \right],
\end{equation}
where $F_{{\rm eff}}$ is the effective quantity, $k(r)=\sqrt{2\mu|Q_{\alpha} - V(r)|}/\hbar$ with $V(r)$ being the
effective potential between the cluster and the daughter nucleus. $R_1$ and $R_2$ are turning points obtained by requiring $V (r) = Q_{\alpha}$.
Similar successfully empirical approaches based on an effective $\alpha$-core potential were also developed in recent publications (Refs. \cite{Xu2006322,0954-3899-26-8-305,PhysRevC.74.014312,Denisov2009815,0954-3899-37-10-105107,PhysRevC.78.057302} and references therein).
An effective $\alpha$-particle equation is derived for the $\alpha$ particle on top of the $^{208}$Pb core in Refs. \cite{PhysRevC.90.034304,	
2015arXiv151107584X}, where an attractive pocket-like potential appears around the nuclear surface. That is related to the sharp disappearance of the nucleon density in
the Thomas-Fermi model employed in their work.

The formation
amplitude $F_{\alpha}(R)$ extracted from the experimental
half-lives data is a model-independent quantity. On the other hand, effective $\alpha$ formation quantities like $F_{{\rm eff}}$ are often introduced in many effective models, which are determined by minimizing the difference between the calculation and the experimental datum. This quantity depends strongly on the shape of the effective potential employed \cite{Buck90}. 
One may wonder how the $\alpha$ formation mechanism is manifested in effective models, which is not explicitly taken into account. 
Since the radius $R$ should
satisfy the relation of $R_1<R<R_2$, we have
\begin{equation}
\Gamma = F_{{\rm eff}}\exp\left[-2\int^{R}_{R_1}k(r)dr
\right]P(R),
\end{equation}
where we define a penetration factor $P$ that, after some mathematics, is given as
\begin{equation}\label{pn}
P=\frac{[H^+_0(\chi,\rho)]^{-2}}{\tan \beta} =
\exp\left[-2\chi(\beta-\sin\beta\cos\beta)\right].
\end{equation}
One thus realizes that the product $F_{{\rm eff}}\exp\left[-2\int^{R}_{R_1}k(r)dr
\right]$ mimics in an effective way the $\alpha$ formation process within the nucleus. By using a properly chosen potential, it is possible to reproduce the general smooth trend of the $\alpha$ formation amplitude. The reduced width introduced in Ref. \cite{PhysRev.113.1593} is also a similar effective quantity that depends on the effective optical potential. 

\subsection{Heavier cluster decays}
The spontaneous emission of clusters heavier than $\alpha$ particle was first observed in 1984 \cite{Nature1984}. It has been
established experimentally in trans-lead nuclei decaying into
daughters around the doubly magic nucleus
$^{208}$Pb. A second island of
cluster radioactivities is expected in trans-tin nuclei decaying
into daughters close to $^{100}$Sn. 

One advantage of the different generalizations of the Geiger-Nuttall law and semiclassical approaches is that, if reliable values of decay $Q$ values can be obtained, it is easy to extrapolate to all kinds of cluster decays
throughout the nuclear chart, which can be a challenging task for microscopic models. 
Systematic calculations on the decays of clusters heavier than $^{4}$He were done in Refs. \cite{Qi2009,QI2009a,0954-3899-39-1-015105,PhysRevC.70.034304}.
Such calculations were extrapolated to the decays of even heavier clusters from superheavy nuclei to daughter nuclei around $^{208}$Pb in Ref. \cite{PhysRevLett.107.062503} and later in Refs. \cite{PhysRevC.85.034615,PhysRevC.89.067301}. However,
further analysis is necessary to understand the uncertainty behind the extrapolation. 

\section{Proton radioactivity}
The proton radioactivity is also shown to be a useful tool to study the structure of nuclei beyond the proton
drip-line. It is often referred to as proton emission or proton decay (not to be confused with the unseen decay of a proton) in nuclear physics. Nearly 50 proton decay events 
have been successfully observed in odd-$Z$ elements between $Z=53$ and $Z=83$ in the past few decades, 
leading to an almost complete
identification of the proton edge of nuclear stability in this region
\cite{doi:10.1146/annurev.nucl.47.1.541,Blank2008a}. 
The concurrence of both proton decay and $\alpha$ decay was also observed  in several nuclei.
On the theoretical side,
the proton-emission process can be described as the quantum 
tunneling of a quasistationary single-particle state  through the Coulomb and centrifugal barriers
 \cite{Delion2006a}. Similar to the case of $\alpha$ decay in Eq. (\ref{foram}), the proton decay formation amplitude can be evaluated as
\begin{equation}
F_l(R)=\int d{\mathbf R} d\xi_d 
[\Psi_d(\xi_d)\xi_pY_l(\mathbf R)]^*_{J_mM_m}
\Psi_m(\xi_d,\xi_p,\mathbf{R}),
\end{equation}
where $d$, $p$ and $m$ label the daughter, proton and mother
nuclei, respectively and $l$ is the orbital angular momentum carried by the outgoing proton. In 
the BCS approach the formation amplitude at a given radius $R$ is
proportional to the product of the occupancy $u$ times the single-proton wave 
function $\psi_p(R)$.  $F_l(R)$ would indeed be the
wave function of the outgoing particle $\psi_p(R)$ if the
mother nucleus behaved simply as
\begin{equation}\label{mother}
\Psi_m(\xi_d,\xi_p,\mathbf{R})=
[\Psi_d(\xi_d)\xi_p\psi_p(R) Y_l(\mathbf R)]_{J_mM_m}.
\end{equation}
One example is the proton-unbound nucleus $^{109}$I \cite{Procter2011118} for which the lowest collective band starting from $7/2^+_1$ and the inner-band E2 transition properties are very similar to those of ground state band in 
$^{108}$Te \cite{PhysRevC.84.041306} and the $7/2^+_1$ band in $^{109}$Te \cite{PhysRevC.86.034308}, indicating that the odd proton in $^{109}$I, which  occupies the $g_{7/2}$ orbital, is  weakly coupled to the $^{108}$Te daughter nucleus like a spectator. This scheme is supported by shell-model and pair truncated shell model calculations \cite{PhysRevC.88.044332}.
The ground state of $^{109}$Te is 98 keV lower than the $7/2_1^+$ state, for which the  spin-parity has been tentatively assigned as $5/2^{+}$. 
This state can be reproduced nicely by the shell model calculation. It is predicted to be dominated by the coupling of a $d_{5/2}$ neutron to $^{108}$Te. Based on systematics of proton decay half-lives \cite{Qi2012c} and the level structure of I isotopes from Ref. \cite{nudat}, a similar $5/2_1^+$ state is also expected to be the ground state of $^{109}$I. However, it was not seen in the life-time measurement in Ref. \cite{Procter2011118}.

The logarithm of the decay half-life can be approximated by \cite{Delion2006,Qi2012c}
 \begin{eqnarray}
\label{gn-3} 
\log T_{1/2}  &=&a\chi' + b\rho' +d l(l+1)/\rho' + c,
\end{eqnarray}
where $a$, $b$, $c$ and $d$ are constants which can be determined by fitting available 
experimental data. It is seen that most of the data can be reproduced by the calculation within a 
factor of four \cite{Qi2012c}.
Relatively large discrepancies are seen for a few
emitters between $63\leq Z\leq 67$ and the isomeric $h_{11/2}$
hole state in  the $Z=81$ nucleus $^{177}$Tl  and the ground state 
of $^{185}$Bi.

\begin{figure}
\begin{center}
\includegraphics[width=0.45\textwidth]{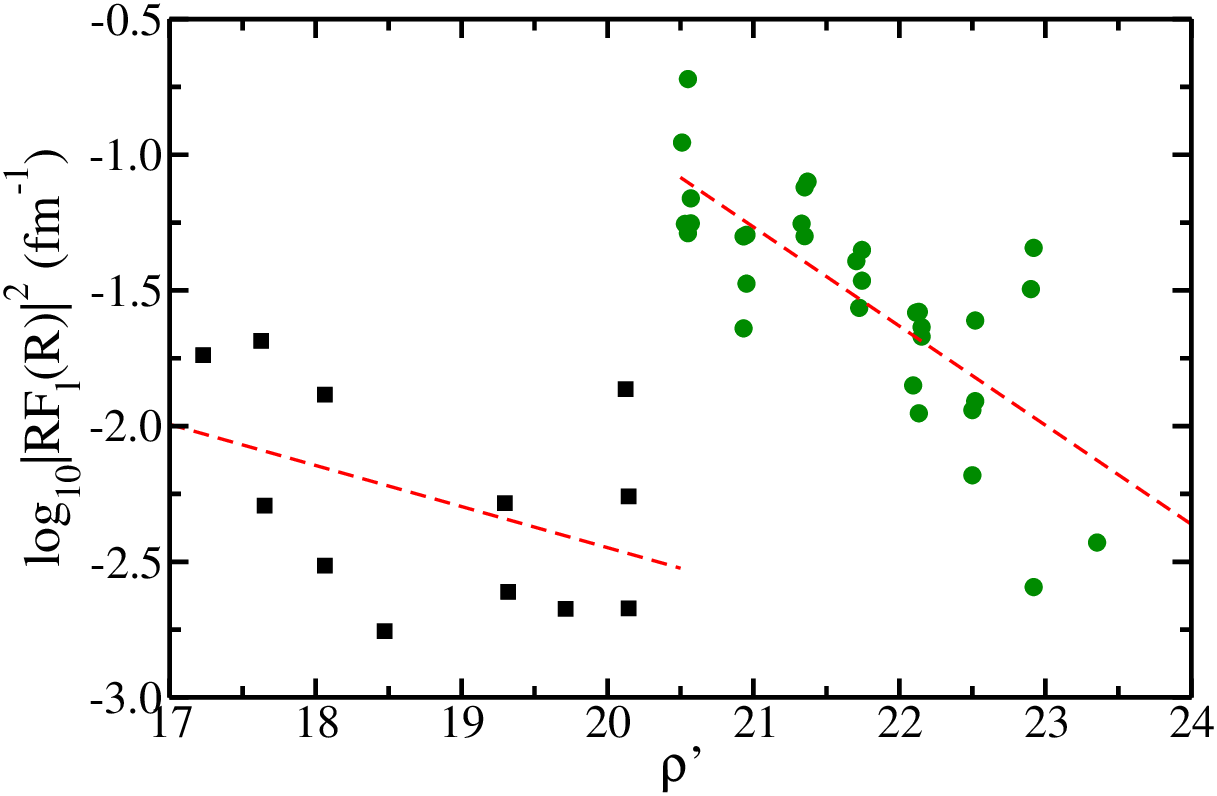}
\end{center}
\vspace{-0.8cm}
\caption{(Color online) The logarithm of the proton-decay formation probabilities
$\log_{10} |RF_l(R)|^{2}$ extracted from experimental data as a function of 
$\rho'$ from \cite{Qi2012c}.  Squares correspond to nuclei with $N<75(Z\leq67)$ while circles 
are for $N\geq 75(Z>67)$.
\label{fvsrho}}
\end{figure}

To further understand Eq. (\ref{gn-3}), the formation amplitudes $F_l(R)$ were extracted from the experimental half lives \cite{Qi2012c}, which are plotted in Fig.~\ref{fvsrho} as a function of $\rho'$. 
One sees that two clearly defined regions emerge. The region to the left in Fig. \ref{fvsrho},
i.e. for lighter isotopes, corresponds to the decays of well deformed nuclei where the formation probabilities decreases for these
nuclei as $\rho'$ increases. Then, suddenly, a strong
transition occurs for the nucleus
$^{144}_{~69}$Tm at $\rho'$=20.5. Here the formation probability acquires its maximum value, where the experimental uncertainty regarding the half-life
(from where the formation probability is extracted) is still quite large,
and then decreases again as $\rho'$ increases. 
The reason of the tendency of the formation probability in the figure is
related to the influence of the deformation: In the left region of Fig. \ref{fvsrho},  
the decays of the deformed nuclei proceed through  
small spherical components of the corresponding deformed orbitals and, 
therefore, the formation probabilities are small.
The right region of Fig. \ref{fvsrho} involves the decays of spherical orbits
as well as major spherical components of deformed orbitals (for example, the
$h_{11/2}$ component of the Nilsson orbital $11/2^- [505]$) which give large proton formation amplitudes.

Another important question is whether the formation probability is
affected by the proton decay  $Q_p$ value. This is not expected since, as shown schematically in Ref. \cite{Qi2012c},
the formation amplitude at the nuclear surface is not sensitive to changes
in the single-particle energy. Neither the BCS amplitudes $u_k$ are much 
affected by the changing of the energy and the potential depth.
On the other hand, the formation amplitude can indeed be sensitive to the nuclear deformation. But this should not be mixed up with the influence of the deformation on the binding energies and the $Q_p$ value.

The  systematic behavior of $Q_p$ values is presented in Ref. \cite{PhysRevC.83.014305} which provides good information for estimating $Q_p$ values for as yet
unknown proton and $\alpha$ decays and for the possibility for them to be observed using current experimental methods. It is suggested that the most likely candidates are $^{158,160}$Re, $^{164,165}$Ir and $^{169}$Au. The partial half-lives for the proton and $\alpha$-decay branches $^{160}$Re are measured to be $687 \pm
11$ $\mu$s and $5.6 \pm0.5$ ms, respectively \cite{PhysRevC.83.064320}. The proton decay is expected to be from the $d_{3/2}$ orbital.
The $\alpha$-decay branch of the
$h_{11/2}$ isomeric state in $^{164}$Ir was identified in Ref. \cite{PhysRevC.89.064309}. 

There have been extensive efforts measuring the rotational
bands of proton emitters
including $^{141}$Ho and the tri-axially deformed nucleus $^{145}$Tm  \cite{PhysRevC.58.R3042,PhysRevLett.86.1458,PhysRevLett.99.082502}.
Moreover, $\gamma$ rays from excited states feeding proton-emitting ground- or isomeric-state  have been observed for $^{112}$Cs \cite{PhysRevC.85.034329}, $^{117}$La \cite{Liu201124}, $^{171}$Au \cite{Back2003}, and $^{151}$Lu \cite{Procter201379,PhysRevC.91.044322}. In the latter case the nucleus was suggested to be of moderate oblate deformation.
A multiparticle spin-trap $19^-$ isomer was discovered in $^{158}$Ta in Ref. \cite{PhysRevLett.112.092501}. The state is unbound to proton decay but shows remarkable stability. Structure calculations have been carried out for those nuclei. In Ref. \cite{PhysRevC.89.014317} the rotational band in $^{141}$Ho is described using the
projected shell model by taking deformed Nilsson quasi-particle orbitals as bases. The $^{145}$Tm is well described as the coupling of of deformed rotational core and the odd proton within the particle-rotor framework in Ref. \cite{PhysRevC.78.041305}.

\section{$\alpha$ decays of $N\sim Z$ nuclei}
The np correlation was neglected  in our calculations for heavy nuclei \cite{Qi2010c} where the two-body clustering is induced by the neutron-neutron (nn) and proton-proton (pp) pair correlations.  This is reasonable since the low-lying neutron and proton single-particle states are very different from each other in those cases and the np correlation is expected to be weak. The $\alpha$ formation amplitude may increase as a result of enhanced isovector (with isospin quantum number $T=1$) nn, pp and np pairing and isoscalar (with $T=0$) np correlation in nuclei with $N\sim Z$ where protons and neutrons occupy the same shells and np correlation is expected to be strong. Therefore, the $\alpha$ decays from $N\sim Z$ nuclei can provide an ideal test ground for our understanding of the np correlation for which there is still  no conclusive evidence after long and extensive  studies (see, recent discussions in Refs. \cite{Ced11,Frauendorf201424,Qi11,Xu2012,Qi2015}).
There has already been a long effort answering the question whether the formation probabilities of neutron-deficient $N\sim Z$ isotopes  are larger compared to those of other nuclei \cite{Seweryniak2006,Liddick2006}.
Moreover, if it is correct, this faster  $\alpha$ decay
would also change the borderline of accessible neutron deficient $\alpha$-decaying nuclei and might be motivation for further experimental work. 
Refs. \cite{Liddick2006} compared the $\alpha$-decay reduced widths for Xe and Te nuclei with that of the textbook $\alpha$-decay isotope $^{212}$Po and neighboring Po isotopes and an enhancement by a factor of 2-3 is seen. 
It was also noticed that  the $|RF_{\alpha}(R)|^2$ value of $^{194}$Rn is larger by a similar factor  compared to the $|RF_{\alpha}(R)|^2$ of $^{212}$Po \cite{Andreyev2013,Qi2014203}.
The $\alpha$ decays of $^{114}$Ba \cite{Mazzocchi200229} and light Xe and Te \cite{Seweryniak2006,Janas2005,Liddick2006} isotopes have also been observed.
The decays of $^{112,113}$Ba as well as $^{108}$Xe and $^{104}$Te may soon be reachable.

\begin{figure}
\begin{center}
\includegraphics[width=0.45\textwidth]{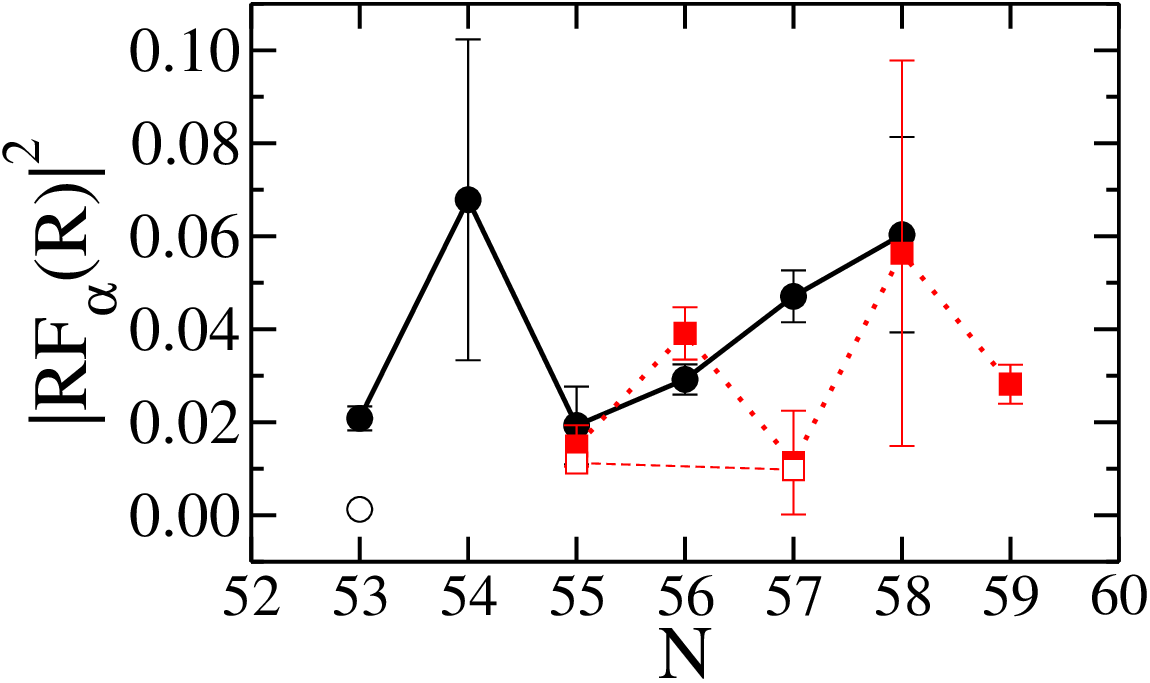}
\end{center}
\vspace{-0.6cm}
\caption{(Color online) $\alpha$-decay formation amplitudes 
$|RF_{\alpha}(R)|^{2}$ extracted from experimental data \cite{nudat,PhysRevLett.105.162502} as a function $N$ for neutron-deficient Te (circle) and Xe (square) above $^{100}$Sn. Open symbols correspond to the decays of $\alpha$ particles carrying orbital angular momentum $l=2$.
\label{fvsrp}}
\end{figure}

In Fig. \ref{fvsrp} we compare the $\alpha$ formation probabilities of nuclei just above $^{100}$Sn. The $\alpha$ formation probabilities of those nuclei follows the general average mass-dependence trend of $\alpha$ formation probability systematics but shows a rather large fluctuations and uncertainties. It is still difficult to determine whether there is indeed an extra enhancement in those transitions. Further experimental investigation
is essential in clarifying the issue. It may be useful to mention here that the systematics of formation probabilities for available $\alpha$ decays shows an increasing trend as the mass number decreases. This is related to the fact that the size of the nucleus also gets smaller, which favors the formation of $\alpha$ particles on the surface.

The influence of np correlation upon the formation of $\alpha$ particles in nuclei $^{220}$Ra and $^{116,108}$Xe was calculated in Ref. \cite{DELION1992407} within the framework of a generalized BCS approach in an axially deformed Woods-Saxon potential. Only diagonal terms between proton and neutron orbitals
with the same angular-momentum projections were considered and a modest enhancement of the clustering was found in $^{116,108}$Xe.
We have evaluated within the shell-model approach the nn and pp two-body clustering in $^{102}$Sn and $^{102}$Te and then evaluated the correlation angle between the two pair by switching on and off the np correlation \cite{Qi2015}. If the np correlation is switched on, in particular if a large number of levels is included, there is significant enhancement of the four-body clustering at zero angle. This is eventually proportional to the $\alpha$ formation probability. It should be mentioned that, one needs large number of orbitals already in heavy nuclei in order to reproduce properly the $\alpha$ clustering at the surface. The inclusion of np correlation will make the problem even more challenging due to the huge dimension.

\section{Summary and outlook}
Understanding how nuclear many-body systems can self-organize in 
simple and regular patterns is a long-standing challenge in modern 
physics. The first case where this was realized concerns  the Geiger-Nuttall law in $\alpha$ 
radioactivity which shows striking linear correlations between the logarithm of 
the decay half-life and the energy of the outgoing particle. We discussed in this review the formation of $\alpha$ particle in nuclei from the clusterization of the two protons and two neutrons through the mixture of high-lying empty single particle configurations, which is induced by the strong pairing correlation. 
We understand that the reason for the success of the Geiger-Nuttall law is
that the $\alpha$-particle formation probability usually varies smoothly from
nucleus to nucleus. Systematics of the $\alpha$ formation probabilities reveal interesting local fluctuations which can provide invaluable information on the pairing correlation and shell structure.
The reduction of the pairing at $Z=82$ and $N=126$ and the changes in the nuclear shapes in neutron-deficient nuclei around $Z=82$ induce suppression of the nuclear $\alpha$ clusterization on the surface. The proton decay can also be an excellent probe for our understanding of the intrinsic structure of the deformed single- (or quasi-) particle orbital.

 It 
will be possible to extend the experimental knowledge on both proton decay and $\alpha$ decay towards more 
neutron deficient nuclei around $Z=82$ and 50 with the new or upgraded radioactive beam facilities. This will allow us to validate the generic description of the $\alpha$ formation probabilities of $N\sim Z$ nuclei above $^{100}$Sn where the influence the np correlation is expected to be the strongest since the protons and neutrons are filling the same single particle orbitals.  It may also shed light on our understanding of np pairing correlation.
More realistic description of the $\alpha$ formation probability in heavy nuclei by using globally optimized density functional and large-scale configuration interaction method may be expected in the near future. A full microscopic description relies also on a realistic choice of the single-particle wave function including the scattering to continuum.
Consequently, more reliable predictions of the $\alpha$ decay half lives will be achieved in unknown nuclei  and in low $\alpha$-decay branching ratios in nuclei close to stability.
\section*{Acknowledgments}
I thank A. N. Andreyev, D.S. Delion, M. Huyse, R. J. Liotta, P. Van Duppen, R. Wyss for collaborations on the present subject. This work was supported by the Swedish Research Council (VR) under grant Nos. 621-2012-3805, 621-2013-4323. I also thank the Swedish National Infrastructure for Computing (SNIC) at NSC in Link\"oping and PDC at KTH, Stockholm for computational support.

\section*{References}
\renewcommand{\bibfont}{\small}
\begin{singlespace}

\setlength{\itemsep}{0pt}
\setlength{\bibsep}{0pt plus 0.3ex}


\providecommand{\newblock}{}

\end{singlespace}

\end{document}